\newcommand{\be}{\begin{eqnarray}}
\newcommand{\ee}{\end{eqnarray}}
\newcommand{\nn}{\nonumber \\}

\newcommand{\CL}{{C\!\ell}}
\newcommand{\p}{\partial}

\newcommand{\openone}{\mbox{1\kern -0.25em I}}
\newcommand{\openK}{\mbox{I\kern -0.25em K}}
\newcommand{\openZ}{\mbox{Z\kern -0.4em Z}}
\newcommand{\openR}{\mbox{I\kern -0.25em R}}
\newcommand{\openH}{\mbox{I\kern -0.25em H}}
\newcommand{\openM}{\mbox{I\kern -0.25em M}}
\newcommand{\openC}{\mbox{C\kern -0.55em I\hspace{0.25em}}}
\newcommand{\con}{\mbox{$\,$\rule{1ex}{0.4pt}\rule{0.4pt}{1ex}$\,$}}
\newcommand{\dwedge}{{\,\dot{\wedge}\,}}
\newcommand{\EOP}{\hfill\rule{4pt}{4pt}}
\newtheorem{dfn}{Definition}
\newtheorem{thrm}[dfn]{Theorem}
\documentstyle[a4]{article}

\begin{document}
\title{On an easy transition from operator dynamics\\
       to generating functionals by Clifford algebras}
\author{Bertfried Fauser\\
Eberhard-Karls-Universit\"{a}t\\
Institut f\"ur Theoretische Physik\\
Auf der Morgenstelle 14\\
72072 T\"ubingen {\bf Germany}\\
Electronic mail: Bertfried.Fauser@uni-tuebingen.de
}
\date{October 24, 1997}
\maketitle
\begin{abstract}
Clifford geometric algebras of multivectors are treated in
detail. These algebras are build over a graded space and exhibit
a grading or multivector structure. The careful study of the
endomorphisms of this space makes it clear, that opposite
Clifford algebras have to be used also. Based on this
mathematics, we give a fully Clifford algebraic account on
generating functionals, which is thereby geometric. The field
operators are shown to be Clifford and opposite Clifford maps.
This picture relying on geometry does not need positivity in
principle. Furthermore, we propose a transition from operator
dynamics to corresponding generating functionals, which is based
on the algebraic techniques. As a calculational benefit, this
transition is considerable short compared to standard ones. The
transition is not injective (unique) and depends additionally on
the choice of an ordering. We obtain a direct and constructive
connection between orderings and the explicit form of the
functional Hamiltonian. These orderings depend on the propagator
of the theory and thus on the ground state. This is invisible in
path integral formulations. The method is demonstrated within
two examples, a non-linear spinor field theory and spinor QED.
Antisymmetrized and normal-ordered functional equations are
derived in both cases.
\end{abstract}

\section{\protect\label{SEC-1}Introduction}

Modern quantum field theory is treated currently by means of
path integrals \cite{GlimmJaff,Rivers}. Path integrals
provide formal solutions of functional differential equations.
Such functional differential equations encode the matrix element
hierarchies of Dyson--Schwinger--Freese
\cite{Dyson,Schwinger,Freese,Lurie}. Generating functionals
describe in a condensed form an infinite set of vacuum or
transition matrix elements with a single mathematical entity.
They prove to be useful in manipulations of the whole hierarchy.

Since path integrals are very compact in their formulation, one
does not actually have an access to their meaning. Moreover, one
is troubled with convergence problems, which can be treated up
to now only by {\it heuristic}\/ methods or in trivial cases,
see chapter 6 of \cite{Rivers}.

It seems therefore to be a good idea, to study generating
functionals and functional differential equations at their own
right. Furthermore, it is known from ordinary differential
equations, that one can achieve informations about the solutions
without being able to integrate the differential equations. We
expect to be able to support the usage of path integrals, at
least their algebraic properties, in studying operator dynamics
with help of functional differential equations.

The paper is organized in three logical parts. First, in section
\ref{SEC-2}, we establish the mathematical apparatus necessary to
treat generating functionals algebraically correct. This will be
done in the case of antisymmetry (fermions), but might be
equally well established for symmetric algebras (bosons), see
\cite{Crumeyrolle}. This part is mathematically rigorous. 

In the second part, consisting of sections \ref{SEC-3} and
\ref{SEC-4}, the infinite dimensional case of QFT is treated
formally, since the transition to infinite many generators
causes serious convergence problems. We give an algebraic
account on generating functionals based on the Clifford algebra
of multivectors. Field operators are shown to be Clifford
endomorphisms of the graded space underlying the Grassmann
algebra build over the Schwinger sources. This opens the first
time an algebraically motivated approach to quantum field
theoretic functionals. In section \ref{SEC-4} we use the
Heisenberg equation to give an very simple transition from
operator dynamics into the space of generating functionals. Once
more, this can be achieved {\it only}\/ if one admits the
Clifford geometric point of view. A single operator dynamics
might result in {\it different}\/ functional equations. This
peculiar observation is an essential insight to understand the
computational power of functional differential equations and
path integrals. The different functional equations obtained are
subjected to different {\it orderings}\/ and ultimately to
different {\it vacua}.\/ The correct {\it choice}\/ of this
ordering by means of normal-ordering with help of a specific
{\it propagator}\/ selects a unique vacuum encoded implicitly
in the theory, as was shown in \cite{Fau-vac}. The Clifford
geometric approach given in this paper makes it possible to
investigate this connections and to have a constructive tool to
implement inequivalent vacua in QFT.

The third part, sections \ref{SEC-5} and \ref{SEC-6} gives
examples of the method. A non-linear spinor field theory,
e.g. a Hubbard or Nambu--Jona-Lasinio type model, and spinor QED as
an example for a boson-fermion coupling theory are treated. Both
examples are dealt with in the antisymmetric and normal-ordered
case. Explicit calculations of the functional equations
shows furthermore the efficiency of our method. The usually
necessary vertex-regularization is no longer apparent, since the
algebraic treatment accounts for this transition correctly. This
was know earlier on the operator level \cite{Fau-ver}. It is
novel, that one can achieve the normal-ordered functional
equation {\it in one single step}.\/

The conclusion summarizes the results and discusses the
relevance to path integral calculations.
 
\section{\protect\label{SEC-2}Clifford geometric algebra of
multivectors} 

There are many possibilities to introduce Clifford algebras,
each of them emphasize a different point of view. In our case,
it is of utmost importance to have the Clifford algebra build
over a graded linear space. This grading is obtained from the
space underlying a Grassmann algebra. The Clifford algebra is
then related to the endomorphism algebra of this space. This
construction, the Chevalley deformation \cite{Che}, was
originally invented to be able to treat Clifford algebras over
fields of $char=2$, see appendix of \cite{MRiesz} by Lounesto
and \cite{Lou}. However, we use this construction in an entire
different context. With help of the construction of M. Riesz
\cite{MRiesz}, one is able to reconstruct the multivector
structure and thereby a correspondence between the linear spaces
underlying the Clifford algebra and the Grassmann algebra in use.
This reconstruction depends on an automorphism $J$, which is
arbitrary, see \cite{Fau-man}. In fact this is just the reversed
direction of our construction following Chevalley given below.

Let $T(V)$ be the tensor algebra build over the $\openK$-linear
space $V$. The field $\openK$ will be either $\openR$ or
$\openC$. With $V^0 \simeq \openK$ we have
\be
T(V) &=& \openK \oplus V \oplus V\otimes_{\openK} V \oplus 
\ldots . 
\ee
The tensor algebra is associative and unital. In $T(V)$ one has
bilateral or two-sided ideals, which can be used to construct
new algebras by factorization. As an example we define the
Grassmann algebra in this was.
\begin{dfn} The Grassmann algebra $\bigwedge(V)$ is the factor
algebra of the tensor algebra w.r.t. the bilateral ideal
\be
I_{Gr} &=& \{ y \mid y=a\otimes x \otimes x \otimes b,\quad
a,b\in T(V),\, x\in V\} \nn
\bigwedge(V) &=& \pi(T(V)) =\frac{T(V)}{I_{Gr}} =
\openK \oplus V \oplus V \wedge V \oplus \ldots .
\ee
The canonical projection $\pi : T(V) \mapsto \bigwedge(V)$ maps
the tensor product $\otimes$ onto the exterior or wedge product
denoted by $\wedge$.\EOP
\end{dfn}
One may note, that the factorization preserves the grading
naturally inherited by the tensor algebra, since the ideal
$I_{Gr}$ is homogeneous. Defining homogeneous parts of $T(V)$
and $\bigwedge(V)$ by $T^k(V)=V\otimes \ldots \otimes V$
and $\bigwedge^k(V)=V\wedge \ldots \wedge V$ $k$-factors, we
obtain $\pi(T^k(V))=\bigwedge^k(V)$.

Proceeding to Clifford algebras requires a further structure,
the quadratic form.
\begin{dfn}\label{dfn-qform}
The map $Q : V \mapsto \openK$, satisfying
\be\label{qform}
i) && Q(\alpha x) = \alpha^2 Q(x),\quad \alpha\in\openK,\, x\in
V \nn
ii) && B_p(x,y) = \frac{1}{2}(Q(x+y)-Q(x)-Q(y)),
\ee
where $B_p(x,y)$ is a symmetric bilinear form is called a
quadratic form.\EOP
\end{dfn}
It is tempting to introduce an ideal $I_{\CL}$
\be\label{idealCL}
I_\CL&=&\{ y \mid y=a\otimes(x\otimes x -Q(x)\openone) \otimes
b, \quad a,b\in T(V),\, x\in V\}
\ee
to obtain the Clifford algebra by a factorization procedure.
However, since we are interested in arbitrary bilinear forms
underlying a Clifford algebra, we will take another approach,
which is wide enough for such a structure. Furthermore, the
Clifford algebra {\it does not}\/ have an intrinsic multivector
structure, but is {\it only}\/ $\openZ_2$ graded, since the
ideal $I_{\CL}$ is inhomogeneous but $\openZ_2$-graded.

Let $V^*$ be the space of linear forms on $V$, i.e. $V^* \simeq
lin[V,\openK]$. Elements $\omega\in V^*$ act on elements $x\in
V$, but there is {\it no natural}\/ connection between $V$ and
$V^*$. However, we can find a set of $x_i$ which span $V$ and
dual elements $\omega_k$ acting on the $x_i$ in a canonical way 
\be
\omega_k(x_i) &=& \delta_{ki}.
\ee
This allows to introduce a map $* : V \mapsto V^*$, $ x_i^* =
\omega_i$ which may be called Euclidean dual isomorphism
\cite{Haft}. The pair $(V^*,V)$ is connected by this duality
which constitutes a pairing $<.\mid .> : V^*\times V \mapsto
\openK$. Since $V^*$ is isomorphic to $V$ in finite dimensions,
it is natural to build a Grassmann algebra $\bigwedge(V^*)$ over
it. This is the algebra of Grassmann multiforms.

It is further a natural thing to extend the pairing of the
grade-one space and its dual to the whole algebras
$\bigwedge(V)$ and $\bigwedge(V^*)$, as can be seen by its
frequent occurrence in literature
\cite{Lou,Fau-man,Ozi-Grass-Cliff,Fau-pos,Fau-diss,AblLou}. This
can be done by the 
\begin{dfn}\label{multiaction}
Let $\tau,\eta \in \bigwedge(V^*)$, $\omega \in V^*$, $u,v \in
\bigwedge(V)$ and $x\in V$, then we can define a canonical
action of $\bigwedge(V^*)$ on $\bigwedge(V)$ requiring
\be\label{tri-rel}
i) && \omega(x) = <\omega \mid x > \nn
ii) && \omega(u\wedge v) = w(u)\wedge v + \hat{u}\wedge
\omega(v) \nn
iii) && (\tau\wedge\eta)(u) = \tau(\eta(u))
\ee
where $\hat{u}$ is the main involution $\hat{V}=-V$ extended to
$\bigwedge(V)$. 
\EOP 
\end{dfn}
In fact we have given by definition \ref{multiaction} an
isomorphism between the Grassmann algebra of multiforms
$\bigwedge(V^*)$ and the dual Grassmann algebra
$[\bigwedge(V)]^*$. This can be made much clearer in writing 
\be&
y\con x = \omega_y(x) = < \omega_y \mid y > = B(y,x),
&\ee
where we have used the canonical identification of $V$ and
$V^*$ via the Euclidean dual isomorphism. One should be very
careful in the distinction of $\bigwedge(V^*)$ and
$[\bigwedge(V)]^*$, since they are isomorphic but not
equivalent. Furthermore, we emphasize that in writing $y\con$ we
make explicitly use of a {\it special}\/ dual isomorphism
encoded in the contraction 
\be
&.\con : V \mapsto V^*& \nn
& y \rightarrow y\con = \omega_y.&
\ee
Since there is no natural, that is mathematically motivated or
even better functorial relation between $V$ and $V^*$, we are
called to seek for {\it physically motivated reasons}\/ to
select a dual isomorphism. 
\begin{thrm}
Let $(V,Q)$ be a pair of a $\openK$-linear space $V$ and $Q$ a
quadratic form as in definition \ref{dfn-qform}. There exists an
injection $\gamma$ called Clifford map from $V$ into the
associative unital algebra $\CL(V,Q)$ which satisfies
\be
\gamma_x \gamma_x &=& Q(x)\openone.
\ee
\EOP
\end{thrm}
\begin{dfn}\label{Cliffalg}
The (smallest) algebra $\CL(V,Q)$ generated by $\openone$ and
$\gamma_x$ is called (the) Clifford algebra of $Q$\/ over $V$.
\EOP
\end{dfn}
By polarization of this relation we get the usual commutation
relations; $x,y\in V$
\be
\gamma_x \gamma_y + \gamma_y\gamma_x &=& 2B_p(x,y),
\ee
where $B_p(x,y)$ is the symmetric polar form of $Q$ as defined
in (\ref{qform}).

\noindent {\bf Remarks:} i) We could have obtained this result
directly factoring the tensor algebra with the ideal
(\ref{idealCL}). ii) There exists Clifford algebras which are
universal, in this case it is convenient to speak from {\it
the}\/ Clifford algebra over $(V,Q)$. iii) If $V\simeq \openK^n
\simeq \openC^n$ or $\openR^n$, we denote $CL(V,Q)$ also by
$CL(\openC^n)\simeq\CL_n$ or $\CL(\openR_{p,q})$ where the pair
$p,q$ enumerate the number of positive and negative eigenvalues
of $Q$. We can as well give the dimension $n$ and signature
$s=p-q$ to classify all quadratic forms over $\openR$. In the
case of the complex field, one remains with the dimension as can
be seen e.g. from the Weyl unitary trick, letting $x_i
\rightarrow i x_i$ which flips the sign. We do not use
sesquilinear forms here, which could be included nevertheless.
\EOP

We will now use Chevalley-deformations to construct the Clifford
algebra of multivectors. The main idea is, that we can decompose
the Clifford map as
\be\label{decompose}
\gamma_x &=& x\con + x\wedge \, .
\ee
We have thus a natural action of $\gamma_x$ on the space
$F\!\bigwedge(V)$.  
\begin{thrm}[Chevalley] 
Let $F\!\bigwedge(V)$ be the space underlying the Grassmann
algebra over $V$ and $\gamma_x$ $:$ $V\mapsto
End(F\!\bigwedge(V))$, $x\in V$ be defined as in
(\ref{decompose}), then $\gamma_x$ is a Clifford map.\EOP 
\end{thrm}
We have shown that $\CL$ is a subalgebra of the endomorphism
algebra of $F\bigwedge(V)$,
\be
\CL &\subseteq& End(F\!\bigwedge(V)).
\ee
It is possible to interpret $x\con$ and $x\wedge$ as
annihilating and creation {\it operators}\/ (on the space
underlying the Grassmann algebra) \cite{Crumeyrolle}.

With help of the relations (\ref{tri-rel}) we can then lift this
Clifford map to multivector actions. {\it No symmetry
requirement has to be made on the contraction.}\/ This leads to
the 
\begin{dfn}[Clifford algebra of multivectors]\label{def-Cliff}
Let $B : V\times V \mapsto \openK$ be an arbitrary bilinear form.
The Clifford algebra $\CL(V,B)$ obtained from lifting the
Clifford map 
\be&
\gamma_x = x\con + x\wedge = <x \mid .> + x\wedge = B(x,.)+x\wedge
&\ee
to $End(F\!\bigwedge(V))$ using the relations (\ref{tri-rel}) is
called Clifford algebra of multivectors. 
\EOP
\end{dfn}
Note, that $B(x,.)=\omega_x$ is a map from $V \rightarrow V^*$
and incorporates a dual isomorphism. It is clear from the
construction that $\CL(V,B)$ has a multivector structure or say
a $\openZ_n$-grading inherited from the graded space
$F\!\bigwedge(V)$.

$B$ admits a decomposition into symmetric and antisymmetric
parts $B=G+F$. The symmetric part $G=B_p$ corresponds to a
quadratic form $Q$, see (\ref{qform}).
\begin{thrm}
The Clifford algebra $\CL(V,Q)\simeq\CL(V,G)$ is isomorphic as
Clifford algebra to $\CL(V,B)$, if $B$ admits a decomposition
$B=G+F$, $G^T=G$, $F^T=-F$.\EOP
\end{thrm}
A proof can be found for low dimensions in \cite{AblLou} and in
general in \cite{Ozi-multivectors}. However, this result was
implicitly known to physicists, see
\cite{Fau-ver,Fau-diss,StuBor}. In fact, this is the old Wick
rule of QFT. We will insist on the $\openZ_n$-grading and
therefore carefully distinguish Clifford algebras of
multivectors with a common quadratic form $Q$ but different
contractions $B$.  

\begin{dfn}
The opposite algebra $A^{op}$ of an algebra $A$ with product
$m(a,b)=ab$ is defined to be the same linear space $F\! A$
underlying $A$ endowed with the opposite or transposed product 
$m^{op}(a,b)=m(b,a)=ba$. 
\EOP 
\end{dfn}
\begin{thrm}[Chevalley]
The opposite Clifford algebra $\CL^{op}(V,Q)$ of $\CL(V,Q)$ is
isomorphic to $\CL(V,-Q)$,
\be\label{thrmClIso}
\CL^{op}(V,Q)&\simeq& \CL(V,-Q).
\ee\EOP
\end{thrm}
We can generalize this theorem to Clifford algebras of
multivectors as
\begin{thrm}
For Clifford algebras of multivectors holds
\be\label{thrmClmult}
\CL^{op}(V,B) &\simeq& \CL(V,-B^T),
\ee
where $T$ denotes the transposition of the bilinear form
$B(x,y)^T=B(y,x)$. 
\EOP
\end{thrm}
The most general linear transformation on $F\!\bigwedge(V)$ is
achieved by left {\it and}\/ right translations, we have 
\be
&
End(F\!\bigwedge V) \simeq
\bigwedge(V)\otimes[\bigwedge(V)]^* \simeq
& \nn &
\CL(V,B) \hat{\otimes} \CL^{op}(V,B)
\simeq
\CL(V,B) \hat{\otimes} \CL(V,-B^T)
\simeq 
& \nn &
CL(V\oplus V,B\oplus -B^T),
&
\ee
see e.g. \cite{Hahn} theorem 5.5 or \cite{BennTuck}. This is the
algebra which will be used in the subsequent sections. 

The fact that in the above construction one has the graded
tensor product is reflected by the following formulas. Let
\be
\gamma_x &=& x\con_B + x\wedge \nn
\gamma^{op}_x &=& x\con_{-B^T} + x\wedge
\ee
be the Clifford maps into $\CL(V,B)$ and $\CL^{op}(V,B)\simeq
\CL(V,-B^T)$, then we obtain the following commutation relations
\be
\gamma_x \gamma_y + \gamma_y \gamma_x ~~~&=& +2G(x,y) \nn
\gamma^{op}_x \gamma^{op}_y + \gamma^{op}_y \gamma^{op}_x \,
&=& -2G(x,y) \nn
\gamma_x \gamma^{op}_y + \gamma^{op}_y \gamma_x ~~&=& 0.
\ee
Remember the decompositions $B=G+F$ and $B^{op}=-B^T=-G+F$. If
one wants to treat ordinary left and right translations
\be
L_ax &=& ax \nn
R_ax &=& xa,
\ee
one has to introduce a further involution which accounts for
the grading in the above graded tensor product of $\CL$ and
$\CL^{op}$. One defines
\be
L_x &=& \gamma_x \nn
R_x &=& \gamma^{op}_x (\hat{.}),
\ee
where $(\hat{.})$ indicates the grade involution {\it
operator}\/ acting to the right. We obtain in this case the
commutation relations as
\be
L_x L_y + L_y L_x =~~~~~ \gamma_x \gamma_y + \gamma_y \gamma_x 
~~~~~~ &=& 2G(x,y) \nn 
R_x R_y + R_y R_x = \gamma^{op}_x \hat{\gamma}^{op}_y(\hat{.}) 
                  + \gamma^{op}_y \hat{\gamma}^{op}_x(\hat{.})
              &=& 2G(x,y) (\hat{.}) \nn
L_x R_y - R_y L_x =\, ~\gamma_x \gamma^{op}_y(\hat{.}) 
                  - \gamma^{op}_y \hat{\gamma}_x(\hat{.}) 
~ &=& 0. 
\ee
We give some further notations. Let $\{j_i\}$ be a set of
elements spanning $V\simeq <j_1,\ldots,j_n >$ and $\{ \p_k\}$ be
a set of dual elements. Building the Grassmann algebras
$\bigwedge(V)$, $\bigwedge(V^*)$ and defining the action of
the forms via (\ref{tri-rel}), one obtains the relations
\be
i)    && j_i \wedge j_i = 0 = \p_i \wedge \p_i \nn
ii)   && \p_i j_k + j_k \p_i = B_{ik}+B_{ki}=2G_{ik}.
\ee
The space ${\bf V}=V\oplus V^T$ is thus spanned by (note the
order of indices)
\be
\{e_1,\ldots,e_{2n} \} &=& 
\{j_1,\ldots,j_n,\p_1,\ldots,\p_n\}.
\ee
Usually, the choice of the contraction is taken to be the
canonical one, with $G_{ik}=\delta_{ik}$. This leads to the
so-called quantum algebra \cite{Saller}. The resulting algebra
is $\CL({\bf V},B^{can})$, where
\be
[B^{can}(e_i,e_j)]= [B^{can}_{ij}]&=&
\begin{array}{|cc|}
0_{n\times n} & \frac{1}{2}\openone_{n\times n} \\
\frac{1}{2}\openone_{n\times n} & 0_{n\times n}
\end{array}.
\ee
This leads to the far more restricted commutation relations of
Sallers quantum algebra (which is in fact a CAR algebra)
\be
i)    && j_i j_k + j_k j_i = 0 = \p_i \p_k + \p_k \p_i \nn
ii)   && \p_i j_k + j_k \p_i = \delta_{ik}.
\ee
However, see \cite{Fau-Hecke} for the usefulness of a more
general bilinear form $B$. We will employ the $j_i$ sources and
$\p_k$ duals to construct field operators. A more general
bilinear form $B=G+F$ is then introduced explicitly via the
{\it quantization} $G$ and the {\it propagator} $F$ in terms of
$j$ sources and $\p$ duals.

\section{\protect\label{SEC-3}Generating functionals}

Generating functionals originated from two major considerations.
The first arose from studying covariant formulated perturbation
theory \cite{Dyson}, see also \cite{Umezawa,Polivanov}, to be
able to handle free-free transitions of particles via a
scattering matrix. The main ingredients are vacuum expectation
values of time-ordered products of field operators. Since in--
and out-going states are assumed to be free, one can retain the
Fock representation, which however is not valid in the
interacting case. Path integrals originated out of this branch
of physics \cite{Fadd80}. 

A second approach was developed to treat QFT in an axiomatic
way \cite{Wigh56}. The main tool are vacuum expectation values
of ordinary operator products. But, beside some structural
results, the most prominent of them the Wightman reconstruction
theorem, the method has not lead to a computational useful tool
\cite{Haag92}, apart from low dimensional models
\cite{GlimmJaff}.  

Both of the constructions assume:
\begin{itemize}
\item[i)] The existence of field operators, and operators build
from them, the hermitean of which constitute the observable of
the theory.
\item[ii)] The existence of a positive, linear, normalized
functional, a state, which allows the calculation of vacuum
expectation values and a statistical interpretation.
\end{itemize}
There are further requirements necessary, but even the given
ones are questionable. It is well known from $C$*-algebra
theory, that representations exist in the thermodynamic limit,
in which no field operators do exist, but only in some larger
closure of the algebra.

We will remain with the first assumption, but tend to weaken the
second one. It is mainly positivity which might be questioned in
this context. QFT deals already with indefinite structures in
describing interactions e.g. Faddeev-Popov ghosts.
Additionally we know, that on orthogonal Clifford algebras,
which are related to fermionic fields, there are no nonsingular
forms in dimensions higher than two (generators)
\cite{Hahn,Knus}. 

It is clear, that a statistical interpretation breaks down
without positivity. But such an interpretation is not called for
in the interaction region. Only in-- and out-going states, which
should have a particle interpretation, have to bear this feature.
This idea was already proposed by Heisenberg.

However, we will go a step further and propose a geometric
notation, which is valid in positive, indefinite and even
singular cases. This makes it possible to decide afterwards in
which situations positivity may be obtained and a statistical
interpretation is meaningful, see \cite{Fau-vac}. The Dirac
theory was already treated within this methods
\cite{Fau-ach}. 

We will treat fermionic quantum fields, but give in section
\ref{SEC-6} also an example for a boson-fermion coupling theory,
which can be developed straight forward.

Our starting point is the {\it quantization rule}\/ imposed on
field operators.
\be\label{Psicom}
\{ \psi_{I_1},\psi_{I_2} \}_+ &=& A_{I_1I_2},
\ee
where $\psi_{I}$ is a `super field' defined as
\be
\psi_I=\psi_{\Lambda K} &=& 
\left\{\begin{array}{cl}
\psi_K & \Lambda=1 \\
\psi^\alpha_K & \Lambda=2.
\end{array}\right. 
\ee
The anti-involution $\alpha$ is taken e.g. to be the hermitean
or charge conjugation. $A_{I_1I_2}$ is a symmetric $c$-number
function of space-time and algebraic freedoms which all together
are summarized in the multiindex $I=\{ \Lambda, \alpha, \ldots
\}$.  

The relation (\ref{Psicom}) can be seen as the defining relation
of a Clifford algebra $\CL(<\psi>,A)$, when we identify the
field operators as generators. To be able to give an algebraic
basis of this Clifford algebra, we need an ordering. Let ${\cal
P}$ denote such an ordering, which acts on the indices of the
field operators. Clifford monomials in the field operators are
then given by
\be
e_n &:=& {\cal P}(\psi_{I_1},\ldots,\psi_{I_n}),
\ee
where ${\cal P}$ might be e.g. time-ordering, antisymmetry for
equal times or normal-ordering w.r.t. a vacuum state. The linear
space underlying  $\CL(<\psi_I >,A)$ is thus the linear span of
the $e_n$ Clifford monomials 
\be
F\! \CL(<\psi_I >,A) &\simeq& <e_n>.
\ee
As we emphasized in section \ref{SEC-2}, it is of utmost
importance to have a graded space underlying those algebras,
which leads to the Clifford algebra of multivectors of
definition \ref{def-Cliff}. In quantum field theory this space
is obtained from an Grassmann algebra by introducing the
anticommuting Schwinger sources $j$ \cite{Schwinger}. Indeed,
due to the relation $j_I^2=0$ this are Grassmann generators.
Assuming that the indices are from the same set as in the case
of the $\psi_I$'s, we build the Grassmann algebra over $V$ with
$V=<j_I>$ 
\be
\bigwedge(<j_I>) &=& \bigwedge(V).
\ee
Of course, we can identify the spaces underlying
$\CL(<\psi_I>,A)$ and $\bigwedge(<j_I>)$ by means of Chevalley's
identification 
\be
F\!\CL(<\psi_I>,A) &\simeq& F\!\bigwedge(<j_I>),
\ee
see section \ref{SEC-2}. This is the graded space physicists
make actually use of in their calculations. This remains true
for path integrals also, as can be seen by their functional
derivatives and the explicit occurrence of Schwinger source terms
$\psi_I j_I$ in the exponent. It is therefore convenient to
relate all entities to this space.

We know from Chevalley deformation, that the Clifford map
$\psi_I$ constitutes an endomorphism of $F\!\bigwedge(<j_I>)$
which does not respects the grading. $\psi_I$ may be decomposed
in a grade lowering or annihilation operator $\p_I$ and a grade
rising or creation operator $1/2 A_{I_1I_2}j_{I_2}\wedge$,
\be\label{psidecomp}
\psi_I &=& \p_I + \frac{1}{2} A_{II_1} j_{I_1}\wedge,
\ee
where we omitted some superfluous factors. The $\p_I$'s are the
canonical duals w.r.t. the Euclidean dual isomorphism. We have
thus the commutation relations
\be\label{com-rel-jp} 
\{ j_I, j_K \}_+ = &0& = \{ \p_I, \p_K \}_+ \nn
\{ j_I, \p_K \}_+ &=& \delta_{IK},
\ee
which results with (\ref{psidecomp}) in the quantization
condition (\ref{Psicom})
\be
\{\psi_I, \psi_K \}_+ &=& A_{IK}.
\ee
The next step towards a QFT generating functional is to project
the Clifford monomials $e_n = {\cal
P}(\psi_{I_1},\ldots,\psi_{I_n})$ onto the graded space
$F\!\bigwedge(<j_I>)$. This is in fact a representation of the
Clifford morphism obtained by lifting the $\psi_I$ action to the
whole algebra, as was described in section \ref{SEC-2}. We
write this projection as $\pi : F\!\CL(V,A) \mapsto
F\!\bigwedge(V)$ 
\be
\pi(e_n) 
&=& \pi({\cal P}(\psi_{I_1},\ldots,\psi_{I_n}))\nn
&=& <\p_{I_n}\wedge\ldots\wedge\p_{I_1}
    {\cal P}(\psi_{I_1},\ldots,\psi_{I_n})>_\pi 
    j_{I_1}\wedge\ldots\wedge j_{I_n} \nn
&=& <0\mid {\cal P}(\psi_{I_1},\ldots,\psi_{I_n}) \mid p>
     j_{I_1}\wedge\ldots\wedge j_{I_n} \nn
&=& \rho_n(I_1,\ldots,I_n\mid p)j_{I_1}\wedge\ldots\wedge j_{I_n}.
\ee
The index $\pi$ parameterizes the linear normed functionals on
$F\!\CL(V,A)$, which is thereafter encoded in the bra $<0\vert$
and ket $\vert p>$ of the notation common to physicists.
Observe, that
\be
<\p_{I_n}\wedge\ldots\wedge\p_{I_1}
    {\cal P}(\psi_{I_1},\ldots,\psi_{I_n})>_\pi 
&=& <0\mid {\cal P}(\psi_{I_1},\ldots,\psi_{I_n}) \mid p>\nn
&=& \rho_n(I_1,\ldots,I_n\mid p)
\ee
is a $c$-number transition matrix element. If $\vert a>$ is set
equal to $\vert 0>$ we obtain vacuum expectation values which
are known to be sufficient to describe the theory.

If a general element of $F\!\bigwedge(<j_I>)$ is written down,
this reads
\be
\vert {\cal P}(j,p)>_F &=& \sum_{i=0}^{n} \frac{i^n}{n!} 
\rho_n(I_1,\ldots,I_n\mid p)j_{I_1}\wedge\ldots\wedge j_{I_n} 
\vert 0>_F\nn
\rho_n(I_1,\ldots,I_n\mid p)&=& <0\mid {\cal
P}(\psi_{I_1},\ldots,\psi_{I_n}) \mid p> 
\ee
and is thus exactly the definition of a fermionic generating
functional in quantum field theory \cite{Lurie,Itzykson}. To
emphasize the `vector'-state character, in fact the left linear
structure, we have added a functional ket $\vert 0>_F$ which is
usually subjected to the relations
\be
{}_F\!<0\mid j_I = &0& = \p_I \mid 0>_F.
\ee
This writings mimic in fact a linear form imposed on the
algebra. We have thus given a Clifford algebraic account on
quantum field theoretic generating functionals. The Clifford
algebra involved was shown to be a Clifford algebra of
multivectors build over a graded space $F\!\bigwedge(<j_I>)$. 

One may note, that due to the explicit introduction of a basis
$\{j_I \}$ the functional is also basis dependent, especially
the matrix elements $\rho_n(I_1,\ldots,I_n\mid p)$ are. However,
from our general considerations it is clear, that the functional
`state' $\vert {\cal P}(j,p)>_F \in \bigwedge(<j_I>)$ is an
invariant object and basis independent.

\section{\protect\label{SEC-4}Transition from operator dynamics
to generating functionals}

The most common case in quantum field theory is that an
operator dynamics is given {\it a priori}.\/ Furthermore,
quantization calls for the Hamiltonian picture which breaks
explicitly covariance \cite{Dir-ham}. This is not a problem as
long as exact calculations are performed, since one is able to
return to the manifest covariant theory.

Given a Hamiltonian, we can use the Heisenberg equation to write
down the dynamics
\be\label{Heis-eq}
i\dot{\psi}_I &=& [\psi, H]_- \nn
H &=& H[\psi].
\ee
H is assumed to generate a one parameter family of
automorphisms by integrating (\ref{Heis-eq}) to
\be
\psi_I(t) = e^{iHt} \psi_I(0) e^{-iHt}.
\ee
It can be shown under the restrictive conditions of $C$*-theory,
that this is always possible. 

We know, that generating functionals are constructed by lifting
the $\psi_I$-action to $F\!\bigwedge(<j_I>)$, so it is a natural
question to ask for a Clifford morphism extension of
(\ref{Heis-eq}). This will be the desired functional equation.

Since it is quite not clear, which $H[\psi]$ integrate to a
group action, we assume this explicitly. However, the space
underlying a Clifford algebra can easily be made into a
Lie-algebra by introducing the commutator product as a
Lie-product. This has not to be confused with the Lie-algebras
obtained as bi-vector sub-algebras under the commutator product.

A straight forward method to obtain derivatives of generating
functionals is the Baker-Campbell-Hausdorff formula. If one writes
the generating functional formally as
\be
e^{\psi(t)} &\simeq& \vert {\cal P}(j,p)>_F
\ee
one has to calculate
\be
i\p_0 e^{\psi(t)} &=& \sum_{k=0}^{\infty} [\psi,H]_k \nn
&& [\psi,H]_k := [[\psi,H]_{k-1},H]_- \quad k \ge 1\nn
&& i\p_0\psi= [\psi,H]_0\,=\,[\psi,H],
\ee
which is quite tedious. Furthermore, one has to apply the BCH
formula a second time to eliminate the field operators, which
prolongates the calculations enormous. The result is then the
functional equation
\be
i\p_0 \vert {\cal P}(j,p)>_F &=& H[j,\p]^{\cal P}
\vert {\cal P}(j,p)>_F
\ee
with the functional Hamiltonian $H[j,\p]^{\cal P}$. This
Hamiltonian depends in its explicit form on the ordering
implicit in $\vert {\cal P}(j,p)>_F$.

We will circumvent this lengthy calculations using the opposite
algebra $\CL^{op}(<\psi_I>,A)$. To account for generating
functionals, we reinterpret the field operators as Clifford
endomorphisms on the space $F\!\bigwedge(<j_I>)$ spanned
by the Grassmann sources. If we allow only a left action, we can
not reach every endomorphism on this space. However, the algebra
$\bigwedge(<j_I>)\otimes \bigwedge(<j_I>^*)$ was shown to be
isomorphic to $End(F\!\bigwedge(<j_I>)$. We have therefore to
construct the opposite Clifford algebra to handle right actions.
In theorem \ref{thrmClIso} we obtained the isomorphism
\be
\CL^{op}(<\psi_I>,A) &\simeq& \CL(<\psi_I>,-A).
\ee
The decomposition into creation and annihilation part reads then
for the antisymmetric ordering
\be
\psi_I &=& \p_I +\frac{1}{2} A_{II^\prime} j_{I^\prime}\wedge \nn
\psi^{op}_I &=& \p_I -\frac{1}{2} A_{II^\prime} j_{I^\prime}\wedge.
\ee
$\psi^{op}$ acts as an operator of right action from the left on
the space $F\!\bigwedge(<j_I>)$. Indeed, one finds
\be
\p_I &=& \frac{1}{2}(\psi_I + \psi^{op}_i) \nn
j_I  &=& A_{II^\prime}(\psi_{I^\prime}-\psi^{op}_{I^\prime}),
\ee
where we assumed $AA=\openone$, according to the Euclidean dual
isomorphism. It is now an easy task to find the desired lifting
of the Clifford morphism.

Let $g_n = \pi(e_n) = \rho_n(I_1,\ldots,I_n\mid p)
j_{I_1}\wedge\ldots\wedge j_{I_n}$ be the projected monomial.
The action of $H[\psi]$ on this monomials $g_n$ may be defined
as 
\be
i\dot{g}_n &=& [g_n, H[\psi] ]_- = g_n H[\psi]-H[\psi] g_n \\
&=& H[\psi^{op}] g_n -H[\psi] g_n.
\ee
A detailed investigation of the grading properties should then
follow. However, to reproduce the results in literature, the
grade involution is suppressed. The action of the Hamiltonian on
multivectors is thus {\it defined}\/ to be 
\be
i\dot{g}_n &=& ( H[\psi^{op}] - H[\psi] ) g_n.
\ee
Summing up the monomials yields the functional equation
\be\label{FKT-Heisenbergeq}
i\p_0 \mid {\cal P}(j,a)>_F &=&  ( H[\psi^{op}] - H[\psi] ) \mid
{\cal P}(j,a)>_F \nn
&=& H[j,\p]^{\cal P} \mid {\cal P}(j,a)>_F,
\ee
where $H[j,\p]^{\cal P}$ is the desired functional Hamiltonian
acting on the whole Dyson-Schwinger-Freese hierarchy encoded in
the generating functional.

Instead of tedious calculations by means of the BCH formula, we
have simply to decompose $\psi$ and $\psi^{op}$ into
$j$-$\p$-parts and sum up the terms. In the above case of an
antisymmetric ordering we have
\be
H[j,\p]^{as} &=& H[\psi^{op}] - H[\psi] \nn
&=& H[\p-1/2\, Aj] - H[\p +1/2\, Aj].
\ee
The antisymmetric ordering was introduced in the second line due
to the {\it special form}\/ of the decomposition of $\psi_I$
into annihilating and creating parts. In the case of an
arbitrary ordering, we have to use the Clifford algebra of
multivectors with an arbitrary bilinear form $B=G+F=A+F$ ($G$
and $A$ are synonyms). The resulting decompositions are (without
factors)
\be
\psi &=& \p +Aj +Fj \nn
\psi^{op} &=& \p -Aj +Fj,
\ee
due to theorem \ref{thrmClmult}. We obtain as $F$-ordered or
say normal-ordered functional Hamiltonian
\be
H[j,\p]^{\cal N} &=& H[\psi^{op}] - H[\psi] \nn
&=& H[\p-Aj+Fj] - H[\p +Aj+Fj]
\ee
and the normal-ordered functional equation
\be
i\p_0 \mid {\cal N}(j,a)>_F &=&  ( H[\psi^{op}] - H[\psi] ) \mid
{\cal N}(j,a)>_F \nn
&=& H[j,\p]^{\cal N} \mid {\cal N}(j,a)>_F.
\ee
The usefulness of this approach will be shown  in the following
sections by explicit calculations. Bosons can be treated in an
analogous way using symplectic Clifford algebras
\cite{Crumeyrolle}. We have not accounted for all factors $i$
and $1/2$ occurring for historical reasons in the definition of
generating functionals, this will be done in the examples.

\section{\protect\label{SEC-5}Non-linear spinor field model}

It will be illuminating to treat every model in two versions.
First the pure antisymmetric case which is the one-time limit of
the time-ordered functionals and afterwards the more general
normal-ordered case, where $F$ is either the correct propagator
of the theory or taken as a {\it parameter}. 

We use a compact notation, where spinor operators and charge
conjugate spinors operators are united
\be
\psi_I = \psi_{\Lambda K} &=&
\left\{ \begin{array}{ll}
\psi_K & \Lambda=1 \\
\psi^c_K & \Lambda=2 .
\end{array}\right.
\ee
It is useful to work with charge conjugate field operators,
since they allow a compact notation of the field equations. We
define our model as (summation and integration over repeated
indices implicitly assumed) 
\be\label{NJL-model}
i\p_0 \psi_I &=& D_{II_1}\psi_{I_1}
        +gV_I^{\{I_1I_2I_2\}_{as}}
        \psi_{I_1}\psi_{I_2}\psi_{I_3} \nn
\{ \psi_{I_1}\psi_{I_2} \}_+ &=& A_{I_1I_2},
\ee
where $D_{II_1}$ is a kinetic term and
$V_I^{\{I_1I_2I_3\}_{as}}$ a constant vertex. The anticommutator
is denoted by $A_{I_1I_2}$ and we require the identity
$A_{I_1I_2}A_{I_2I_3}=\delta_{I_1I_3}$. For explicit
expressions see e.g. (\ref{SQED-def}). But, it is our intention
not to give explicit expressions for $D,V$ and $A$, since we
want to demonstrate the model independency of our method.

The antisymmetry of $V_I^{\{I_1I_2I_2\}_{as}}$ can be pushed
over to the field operator products. In this case one has to be
careful since there are relations of the type
\be
:\,\psi_{I_2}\wedge\psi_{I_3}\wedge\psi_{I_4}\,:\,\, &=&
\psi_{I_2}\dwedge\psi_{I_3}\dwedge\psi_{I_4}\nn
&=&\psi_{I_2}\wedge\psi_{I_3}\wedge\psi_{I_4}
+F_{I_2I_3}\psi_{I_4}-F_{I_2I_4}\psi_{I_3}+F_{I_3I_4}\psi_{I_2},
\ee
which result from the normal-ordering \cite{Fau-ver}. We have
thus carefully to distinguish the different wedge products,
which ultimately lead to different vacua \cite{Fau-vac}.
Literally this is done by the $:\ldots :$ notations used by
physicists, but there without an algebraic theory behind. From
the dynamics (\ref{NJL-model}) we obtain the Hamilton operators
\be\label{H-keil}
H[\psi]^\wedge
&=&\frac{1}{2}A_{I_1I_3}D_{I_3I_2}\psi_{I_1}{\wedge}\psi_{I_2} 
+\frac{g}{4}A_{I_1I_5}V_{I_1}^{I_2I_3I_4}
\psi_{I_1}{\wedge}\psi_{I_2}{\wedge}\psi_{I_3}{\wedge}\psi_{I_4}
\ee
and
\be\label{H-dkeil}
H[\psi]^{\dwedge}
&=&\frac{1}{2}A_{I_1I_3}D_{I_3I_2}\psi_{I_1}\dwedge\psi_{I_2}
+\frac{g}{4}A_{I_1I_5}V_{I_1}^{I_2I_3I_4}
\psi_{I_1}\dwedge\psi_{I_2}\dwedge\psi_{I_3}\dwedge\psi_{I_4},
\ee
where the superscript wedges indicate the usage of the wedge or
doted wedge products. 

\subsection{\protect\label{SEC-5.1}Antisymmetic case}

The antisymmetric case is connected with the normal wedge.
Furthermore, the field operators corresponding to this ordering
are given as
\be
\psi &=&\frac{1}{i} \p_i -\frac{i}{2}A_{II_1}j_{I_1}\wedge
\quad \simeq \quad \p-Aj \nn
\psi^{op} &=&\frac{1}{i} \p_i +\frac{i}{2}A_{II_1}j_{I_1}\wedge 
\quad \simeq \quad \p+Aj. 
\ee
Using the functional Heisenberg equation
(\ref{FKT-Heisenbergeq}), we have to calculate
\be
i\partial_0\vert{\cal A}(j,a)>_F&=&E_{0a}\vert{\cal
A}(j,a)>_F\nn 
&=&
(H[\partial-Aj]-H[\partial+Aj])\vert{\cal A}(j,a)>_F,
\ee
where $\vert a>$ is assumed to be an energy eigen-`state' here
and in the sequel. Hence, we have to decompose the Clifford map
$\psi$ and $\psi^{op}$ into creation and
annihilation parts w.r.t. the space $F\!\bigwedge(<j_I>)$. The
first term of (\ref{H-keil}) is given as
\be
T_1[\psi]^\wedge
&=&\frac{1}{2}A_{I_1I_3}D_{I_3I_2}\psi_{I_1}\wedge\psi_{I_2},
\ee
which results in functional form as
\be
T_1[j,\partial]^\wedge
&=&T_1[\partial+Aj]-T_1[\partial-Aj]\nn
&=&
\frac{1}{2}A_{I_1I_3}D_{I_3I_2}\Big(\nn
&&
(\frac{1}{i}\partial_{I_2}+\frac{i}{2}A_{I_2I_2^\prime}j_{I_2^\prime})
(\frac{1}{i}\partial_{I_1}+\frac{i}{2}A_{I_1I_1^\prime}j_{I_1^\prime})\nn
&&-
(\frac{1}{i}\partial_{I_2}-\frac{i}{2}A_{I_2I_2^\prime}j_{I_2^\prime})
(\frac{1}{i}\partial_{I_1}-\frac{i}{2}A_{I_1I_1^\prime}j_{I_1^\prime})
\Big)\nn
&=&
\frac{1}{2}A_{I_1I_3}D_{I_3I_2}\Big(\nn
&&
+\frac{1}{i^2}\{\partial_{I_2},\partial_{I_1} \}_+ 
+\frac{i^2}{4}A_{I_2I_2^\prime}A_{I_1I_1^\prime}
\{ j_{I_2^\prime},j_{I_1^\prime} \}_+\nn
&&
+4\frac{i}{2i}A_{I_2I_2^\prime}j_{I_2^\prime}\partial_{I_1}\Big).
\ee
Using the commutation relations of the $j,\p$ generators and the
antisymmetry of $AD$, we obtain
\be
T_1[j,\partial]^\wedge
&=&
-\frac{1}{2}A_{I_2I_3}D_{I_3I_1}
2A_{I_2I_2^\prime}j_{I_2^\prime}\partial_{I_1}\nn
&=&
-D_{I_2^\prime I_1}j_{I_2^\prime}\partial_{I_1}\nn
&=&
-D_{I_1I_2}j_{I_1}\partial_{I_2}.
\ee
The interaction term
\be
T_2[\psi]^\wedge
&=&
\frac{g}{4}A_{I_1I_5}V_{I_5}^{I_2I_3I_4}
\psi_{I_1}\wedge\psi_{I_2}\wedge\psi_{I_3}\wedge\psi_{I_4}
\ee
is decomposed as
\be
T_2[j,\partial]^\wedge
&=&T_2[\partial+Aj]-T_2[\partial-Aj]\nn
&=&
\frac{g}{4}A_{I_1I_5}V_{I_5}^{I_2I_3I_4}\Big\{\times\nn
&&\times
(\frac{1}{i}\partial_{I_4}+\frac{i}{2}A_{I_4I_4^\prime}j_{I_4^\prime})
(\frac{1}{i}\partial_{I_3}+\frac{i}{2}A_{I_3I_3^\prime}j_{I_3^\prime})
\times\nn
&&
\times
(\frac{1}{i}\partial_{I_2}+\frac{i}{2}A_{I_2I_2^\prime}j_{I_2^\prime})
(\frac{1}{i}\partial_{I_1}+\frac{i}{2}A_{I_1I_1^\prime}j_{I_1^\prime})
\nn
&&-
(\frac{1}{i}\partial_{I_4}-\frac{i}{2}A_{I_4I_4^\prime}j_{I_4^\prime})
(\frac{1}{i}\partial_{I_3}-\frac{i}{2}A_{I_3I_3^\prime}j_{I_3^\prime})
\times\nn
&&
\times
(\frac{1}{i}\partial_{I_2}-\frac{i}{2}A_{I_2I_2^\prime}j_{I_2^\prime})
(\frac{1}{i}\partial_{I_1}-\frac{i}{2}A_{I_1I_1^\prime}j_{I_1^\prime})
\Big\}.
\ee
Since $T_2[j,\p]^\wedge$ is an even function of the generators,
we remain with terms of an odd number of $j$ and $\p$ generators.
Furthermore we use the binomial theorem. This can be done, since
new commutator terms does not arise due to antisymmetry of $AV$ in
$I_1,\ldots,I_4$. Accounting finally for a factor $2$ from the
two commutator terms, the result is 
\be
T_2[j,\partial]^\wedge
&=&
gA_{I_4I_5}V_{I_5}^{I_1I_2I_3}
A_{I_4I_4^\prime}j_{I_4^\prime}
\partial_{I_3}\partial_{I_2}\partial_{I_1}\nn
&&+
\frac{g}{4}A_{I_4I_5}V_{I_5}^{I_1I_2I_3}
A_{I_4I_4^\prime}j_{I_4^\prime}
A_{I_3I_3^\prime}j_{I_3^\prime}
A_{I_2I_2^\prime}j_{I_2^\prime}\partial_{I_1}\nn
&=&gj_{I_1}V_{I_1}^{I_2I_3I_4}\Big(
\partial_{I_4}\partial_{I_3}\partial_{I_2}
+\frac{1}{4}
A_{I_4I_4^\prime}A_{I_3I_3^\prime}
j_{I_4^\prime}j_{I_3^\prime}\partial_{I_2}\Big).
\ee
The functional Hamiltonian $H[j,\p]^\wedge$ is the sum of
$T_1^\wedge$ and $T_2^\wedge$. Therefore, the functional energy
equation reads
\be\label{Fk-expl-1}
E_{a0}\vert{\cal A}(j,a)>_F&=&
\Big\{D_{I_1I_2}j_{I_1}\partial_{I_2}\\
&&
+
gj_{I_1}V_{I_1}^{I_2I_3I_4}\Big(
\partial_{I_3}\partial_{I_2}\partial_{I_1}
+\frac{1}{4}
A_{I_4I_4^\prime}A_{I_3I_3^\prime}
j_{I_4^\prime}j_{I_3^\prime}\partial_{I_2}\Big)\Big\}
\vert{\cal A}(j,a)>_F \nonumber
\ee
in accordance with formula (3.98) in \cite{StuBor}.

\subsection{\protect\label{SEC-5.2}Normal-ordered case}

The normal-ordered functional equation is related to the doted
wedge. It contains explicitly the propagator. This results in
another functional Hamiltonian $H[j,\p]^{\dwedge}$. The
functional energy equation reads
\be
i\partial_0\vert {\cal N}(j,a)>_F&=&
E_{a0}\vert {\cal N}(j,a)>_F\nn
&=&\Big(H[\psi^{op}]
       -H[\psi]\Big)
\vert {\cal N}(j,a)>_F,
\ee
where
\be
\psi^{op} &=& \frac{1}{i}\p + \frac{i}{2} Aj\dwedge =
\frac{1}{i}\p+\frac{i}{2}Aj\wedge+iFj\wedge 
\quad\simeq\quad \p +Aj+Fj \nn
\psi      &=& \frac{1}{i}\p - \frac{i}{2} Aj\dwedge =
\frac{1}{i}\p-\frac{i}{2}Aj\wedge+iFj\wedge
\quad\simeq\quad \p -Aj+Fj
\ee
are the Clifford maps related to the doted wedge, written in
terms of the undoted ones. This allows us to use the same $j$
and $\p$ generators. Since the operator dynamics is formally the
same, we have once more to calculate $T_1$ and $T_2$. Now
\be
T_1[j,\partial]^{\dwedge}
&=&T_1[\partial+Aj+Fj]-T_1[\partial-Aj+Fj]\nn
&=&
\frac{1}{2}A_{I_1I_3}D_{I_3I_2}\Big\{\times\nn
&&\hspace{-2cm}
~~(\frac{1}{i}\partial_{I_2}
+\frac{i}{2}A_{I_2I_2^\prime}j_{I_2^\prime}
+iF_{I_2I_2^\prime}j_{I_2^\prime})
(\frac{1}{i}\partial_{I_1}
+\frac{i}{2}A_{I_1I_1^\prime}j_{I_1^\prime}
+iF_{I_1I_1^\prime}j_{I_1^\prime})\nn
&&\hspace{-2cm}
-
(\frac{1}{i}\partial_{I_2}
-\frac{i}{2}A_{I_2I_2^\prime}j_{I_2^\prime}
+iF_{I_2I_2^\prime}j_{I_2^\prime})
(\frac{1}{i}\partial_{I_1}
-\frac{i}{2}A_{I_1I_1^\prime}j_{I_1^\prime}
+iF_{I_1I_1^\prime}j_{I_1^\prime})\Big\}
\ee
is no longer odd or even in $j,\p$ and we can only use
antisymmetry arguments. Inspection of the equations shows, that
only odd numbers of $A$'s can occur. This are
\be
Aj\partial &:& 2\frac{i}{2i} 
A_{I_2I_2^\prime}j_{I_2^\prime}\partial_{I_1}\nn
AFjj &:& 2\frac{i^2}{2}
A_{I_2I_2^\prime}F_{I_1I_1^\prime}
j_{I_2^\prime}j_{I_1^\prime}.
\ee
With the same considerations as above, we obtain
\be
T_1[j,\partial]^{\dwedge}
&=&
-D_{I_1I_2}j_{I_1}\partial_{I_2}
+
D_{I_1I_3}F_{I_3I_2}j_{I_1}j_{I_2}.
\ee
The interaction term $T_2$ is treated along the same lines. We
have to use the polynomial theorem and the polynomial
coefficients 
\be
P_{k_1\ldots k_n}:=\frac{k!}{k_1!\ldots k_n!},\quad k=\sum k_n .
\ee
The resulting $6$ terms occurring twice are given in table 1.
\medskip

\begin{tabular}{cccl}
\hline\hline
\multicolumn{4}{c}{\bf Table 1.}\\
\hline\hline
($\#\partial, \# Aj, \#Fj$) & $P_{k_1\ldots k_n}$ & factor &
resulting term\\
\hline
(3,1,0) & $ \frac{4!}{3!}=4$ & $4\frac{i}{2i^3}=-2$ & 
$-2A_{I_4I_4^\prime}j_{I_4^\prime}
\partial_{I_3}\partial_{I_2}\partial_{I_1}$
\\
(2,1,1) & $\frac{4!}{2!}=12$ & $12\frac{i^2}{2i^2}= 6$ &
$6 A_{I_4I_4^\prime}F_{I_3I_3^\prime}
j_{I_4^\prime}j_{I_3^\prime}
\partial_{I_2}\partial_{I_1}$
\\
(1,3,0) & $\frac{4!}{3!}=4$ & $4\frac{i^3}{2^3i}= -\frac{1}{2}$ &
$-\frac{1}{2}
A_{I_4I_4^\prime}A_{I_3I_3^\prime}A_{I_2I_2^\prime}
j_{I_4^\prime}j_{I_3^\prime}j_{I_2^\prime}
\partial_{I_1}$
\\
(1,1,2) & $\frac{4!}{2!}=12$ & $12\frac{i^3}{2i}= -6$ &
$-6 A_{I_4I_4^\prime}F_{I_3I_3^\prime}F_{I_2I_2^\prime}
j_{I_4^\prime}j_{I_3^\prime}j_{I_2^\prime}
\partial_{I_1}$
\\
(0,3,1) & $\frac{4!}{3!}=4$ & $4\frac{i^4}{2^3}= \frac{1}{2}$ &
$\frac{1}{2}
A_{I_4I_4^\prime}A_{I_3I_3^\prime}A_{I_2I_2^\prime}F_{I_1I_1^\prime}
j_{I_4^\prime}j_{I_3^\prime}j_{I_2^\prime}j_{I_1^\prime}$
\\
(0,1,3) & $\frac{4!}{3!}=4$ & $4\frac{ii^3}{2}= 2$ &
$2 A_{I_4I_4^\prime}F_{I_3I_3^\prime}F_{I_2I_2^\prime}
F_{I_1I_1^\prime} j_{I_4^\prime}j_{I_3^\prime}j_{I_2^\prime}
j_{I_1^\prime}$.\\
\hline
\end{tabular}
\medskip

After juggling the indices and summarizing we obtain
\be
T_2[j,\partial]^{\dwedge}
&=&
gV_{I_1}^{I_2I_3I_4}\Big[
-j_{I_1}\partial_{I_4}\partial_{I_3}\partial_{I_2}
+3F_{I_2I_2^\prime}j_{I_1}j_{I_2^\prime}
\partial_{I_3}\partial_{I_4}\nn
&&
-(3F_{I_3I_3^\prime}F_{I_2I_2^\prime}
  +\frac{1}{4}A_{I_3I_3^\prime}A_{I_2I_2^\prime})
  j_{I_1}j_{I_3^\prime}j_{I_2^\prime}
  \partial_{I_4}\nn
&&
-(3F_{I_3I_3^\prime}F_{I_2I_2^\prime}
  +\frac{1}{4}A_{I_3I_3^\prime}A_{I_2I_2^\prime})
  F_{I_1I_1^\prime}
  j_{I_1}j_{I_3^\prime}j_{I_2^\prime}j_{I_1^\prime}\Big].
\ee
The normal-ordered functional energy equation obtained {\it
directly}\/ from the operator dynamics in one single step reads
then 
\be
E_{a0}\vert {\cal N}(j,a)>_F
&=&
H[j,\partial]^{\dwedge} \vert {\cal N}(j,a)>_F\nn
&=&\Big\{
D_{I_1I_2}j_{I_1}\partial_{I_2}
-D_{I_1I_3}F_{I_3I_2}j_{I_1}j_{I_2}\nn
&&
+gV_{I_1}^{I_2I_3I_4}\Big[
 j_{I_1^\prime}\partial_{I_4}\partial_{I_3}\partial_{I_2}
-3F_{I_2I_2^\prime}j_{I_1}j_{I_2^\prime}
\partial_{I_3}\partial_{I_4}\nn
&&
+(3F_{I_3I_3^\prime}F_{I_2I_2^\prime}
  +\frac{1}{4}A_{I_3I_3^\prime}A_{I_2I_2^\prime})
  j_{I_1}j_{I_3^\prime}j_{I_2^\prime}
  \partial_{I_4}\\
&&
+(3F_{I_3I_3^\prime}F_{I_2I_2^\prime}
  +\frac{1}{4}A_{I_3I_3^\prime}A_{I_2I_2^\prime})
  F_{I_1I_1^\prime}
  j_{I_1}j_{I_3^\prime}j_{I_2^\prime}j_{I_1^\prime}\Big]
\vert {\cal N}(j,a)>_F.\nonumber 
\ee
$H[j,\p]^{\dwedge}$ is the normal-ordered functional
Hamiltonian. The result agrees with (3.106) in \cite{StuBor}. 

{\bf Remark:} In QFT the normal-ordered functional equation is
obtained in two steps. Only after deriving the antisymmetric
functional equation normal-ordering is performed explicitly via
the non-perturbative Wick transformation
\be
\vert {\cal A}(j,a)>_F &=& e^{-1/2\, jFj}\vert {\cal
N}(j,a)>_F\nn 
H[j,\p]^{\dwedge} &=& e^{-1/2\, jFj} H[j,\p]^\wedge e^{+1/2\,
jFj}.
\ee
Observe, that $j$ commutes with $exp\,({-1/2\, jFj})$, but not
$\p$, so that the transition can be given in terms of generators
also 
\be
\p e^{-1/2\, jFj} &=& e^{-1/2\, jFj} (\p-Fj) \nn
\mbox{d} &:=& \p -Fj \nn
H[j,\p]^{\dwedge} &=& H[j,\mbox{d}]^\wedge,
\ee
compare \cite{Fau-ach}. This method requires explicit vertex
regularization, if the products $\wedge$ and $\dwedge$ are not
explicitly treated as distinct $j_{I_1} \dwedge j_{I_2} = j_{I_1}
\wedge j_{I_2}  +F_{I_1I_2}$, as was shown in \cite{Fau-ver}.

\section{\protect\label{SEC-6}Spinor quantum electrodynamics}

We give spinor QED as an example for a boson-fermion coupling
theory. We have to deal with a constrained theory, since spinor
QED is a gauge theory. Especially on the quantum level this is a
difficult task \cite{StuPfi-i}. Hence in a first step we will
eliminate the gauge freedom classically and perform the
quantization afterwards. This is only possible in QED, because
of the simple Abelian gauge coupling. In non abelian Yang--Mills
Theories as QCD one has a nonlinear term in the Gauss' law, and
a classical elimination leads to difficulties. However, this
can be done using functional techniques
\cite{StuPfi-i,StuPfi-ii}. 

We start with the equations of classical spinor ED
\index{Spinor electrodynamics, covariant}
\be\label{qed-constr}
i)  &&\partial_\nu F^{\mu\nu}+\frac{ie_0}{2}
      \Psi C\gamma^{\mu}\sigma^2\Psi=0 \nn
ii) && (i\gamma^\mu\partial_\mu-m_0)\Psi
       +e_0 A_\mu\gamma^\mu\sigma^3\Psi=0\nn
iii)&&F^{\mu\nu}:=\partial^\mu A^\nu-\partial^\nu A^\mu,
\ee
where we have used the definition \index{Definition, of
super--index} 
\be
\Psi &=& \Psi_{\alpha\Lambda}=\left\{
\begin{array}{ll}
\Psi_\alpha & \Lambda=1 \\
\Psi^c_\alpha=C_{\alpha\beta}\bar{\Psi}_\beta & \Lambda=2
\end{array}\right.
\ee
and omit the matrix indices. By definition the sigma matrices
act on the super index $\Lambda$ distinguishing the
field and its charge conjugated field, which has to be treated
as an independent quantity. The gamma matrices and the 
charge conjugation matrix $C$ act on the spinor index $\alpha$
of the $\Psi$ field.

We use the Coulomb gauge and set  
\be
\partial_k A^k=0.
\ee
Furthermore we split the `bosonic' fields $A$ and $E$ into
longitudinal and transverse parts.
\be
A_k\equiv A^{tr}_k,   && \mbox{because of}\quad\partial_k A^k=0 \nn
E_k=E^{tr}_k+E^{l}_k, && E_k:=-F^{0k}=F_{0k}.
\ee
Furthermore we decompose the field equations into dynamic ones
and the remaining constraint, the Gauss' law. \index{Gauss' law}
Looking at (\ref{qed-constr}-i) yields for
\be\label{Gauss}
\nu=0: && i\partial_0 E_k=-\frac{1}{2}e_0\Psi C\gamma^k\sigma^2\Psi
         +i(\partial_j\partial^k A^j-\partial_j\partial^j A^k)\nn
\nu=k: && \partial^k E_k=-i\frac{e_0}{2}\Psi C\gamma_0 \sigma^2\Psi
\ee
and from the definition of $E^k$
\be
i\partial_0 A^{tr,k}=iE^k-i\partial_k A^0.
\ee
With help of the current conservation the Gauss' law
(\ref{Gauss}-$\nu=k$) can now be used  to yield an expression
for the $A_0$ field in terms of spinors
\be
A_0=-i\frac{e_0}{2}\Delta^{-1}\Psi C\gamma_0\sigma^2\Psi.
\ee
The $\Delta^{-1}$ symbol is used for the integral kernel of the
inverse of the Laplacian. For a further compactification of the
notation we introduce a `super field' $B_\eta$ in the `bosonic'
fields also,
\be
B_k^\eta&:=& \left\{\begin{array}{cl}
A^{tr}_k & \eta=1\\
E^{tr}_k & \eta=2\end{array}\right. .
\ee
The equations of motion reads then
\be\label{dyn}
i\partial_0\Psi_{I_1}&=&D_{I_1I_2}\Psi_{I_2}+W^k_{I_1I_2}B_k\Psi_{I_2}
                 +U_{I_1}^{I_2I_3I_4}\Psi_{I_2}\Psi_{I_3}\Psi_{I_4}\nn
i\partial B_{K_1}&=&L_{K_1K_2}B_{K_2}+J_{K_1}^{I_1I_2}\Psi_{I_1}\Psi_{I_2}
\ee
with the definitions
\be\label{SQED-def}
I                  &:=& \{\alpha,\Lambda,{\bf r}\}\nn
K                  &:=& \{k,\eta,{\bf z}\}\nn
P^{tr}             &:=& 1-\Delta^{-1}\nabla\otimes\nabla\nn
D_{I_1I_2}         &:=&-(i\gamma_0\gamma^k\partial_k-\gamma_0
                       m)_{\alpha_1\alpha_2}\delta_{\Lambda_1\Lambda_2}
                       \delta ({\bf r}_1-{\bf r}_2)\nn
W^K_{I_1I_2}       &:=&e_0(\gamma_0\gamma^k)_{\alpha_1\alpha_2}
                       \delta({\bf r}_1-{\bf r}_2)
                       \delta({\bf r}_1-{\bf z})\delta_{1\eta}
                       \sigma^3_{\Lambda_1\Lambda_2}\nn
U_{I_1}^{I_2I_3I_4}&:=&-\frac{i}{8\pi}e_0^2\left[
                       (C\gamma_0)_{\alpha_2\alpha_3}
                       \delta_{\alpha_1\alpha_4}
                       \sigma^2_{\Lambda_2\Lambda_3}
                       \sigma^2_{\Lambda_1\Lambda_4}
                       \frac{\delta({\bf r}_2-{\bf r}_3)
                       \delta({\bf r}_1-{\bf r}_4)}{
                       \vert{\bf r}_1-{\bf r}_2\vert}\right]_{
                       as\{I_2I_3I_4\}} \nn
L_{K_1K_2}         &:=&i\delta({\bf z}_2-{\bf z}_1)\delta_{k_1k_2}
                       \delta_{\eta_11}\delta_{\eta_22}
                       +i\Delta({\bf z}_1)\delta({\bf z}_1-{\bf z}_2)
                       \delta_{k_1k_2}\delta_{\eta_12}
                       \delta_{\eta_21}\nn
J_K^{I_1I_2}       &:=&-\frac{1}{2}e_0 P^{tr}({\bf z}-{\bf r}_1)
                       \delta({\bf r}_1-{\bf r}_2)(C\gamma^k)_{
                       \alpha_1\alpha_2}\delta_{2\eta}
                       \sigma^2_{\Lambda_1\Lambda_2}.
\ee
We impose commutation relations on the $\Psi$ and $B$ fields
which become thereby operators acting on a suitable state space.
These states should belong to an Hilbert space, if the theory
would be renormalized. The renormalization was discussed in
\cite{StuFauPfi} in a perturbative manner. As a postulate we
introduce the equal time commutation relations 
\be\label{comm-rel}
i)  && \{\Psi_{I_1},\Psi_{I_2}\}^t_+:=A_{I_1I_2}=C\gamma_0\sigma^1
       \delta({\bf r}_1-{\bf r}_2)\nn
ii) && [B_K,\Psi_{I}]^t_-:=0\nn
iii)&& [B_{K_1},B_{K_2}]_-^t=:C_{K_1K_2}.
\ee
Equation (\ref{comm-rel}-i) is the canonical commutation
relation for $\Psi_\alpha$ and $\bar{\Psi}_\alpha$, the Dirac
adjoint field. In our notation with charge conjugate spinor
operators, the somehow unusual $A_{I_1I_2}$ occurs. Because of
the super field notation we have all four commutators belonging
to the $\Psi_\alpha$ and $\bar{\Psi}_\alpha$ integrated in this
relation. 

Equation (\ref{comm-rel}-ii) states, that the boson fields are
looked at as elementary fields. $B_K$ is {\it not} a function of
the $\Psi_I$ fields. Equation (\ref{comm-rel}-ii) is usually
postulated in QED.

Relation (\ref{comm-rel}-iii) specifies the boson commutator.
The special form of the c-number function $C_{K_1K_2}$ can be
calculated as a consequence of requiring a consistent quantum
field theory \cite{StuFauPfi}, see below (\ref{cons}). 

The set of equations (\ref{dyn}) and (\ref{comm-rel}) are the
defining relations of Coulomb gauged quantized spinor ED. All
previous steps are classical and only for convenience to state a
somehow selfconsistent QFT. If one would quantize first and then
try to eliminate the Gauss' law, one would yield an other QFT!
The functional Hamiltonians (see below) would then have infinite
many terms \cite{Lee-QED}.

We obtain the Hamiltonian corresponding to the equations
(\ref{dyn}) as
\be\label{HPsiB}
H(\Psi ,B)&=&\frac{1}{2}A_{I_1I_3}D_{I_3I_2}\Psi_{I_1}\Psi_{I_2}
            +\frac{1}{2}A_{I_1I_3}W^K_{I_3I_2}B_K\Psi_{I_1}\Psi_{I_2}\nn
          &&+\frac{1}{4}A_{I_1I_5}U_{I_5}^{I_2I_3I_4}
             \Psi_{I_1}\Psi_{I_2}\Psi_{I_3}\Psi_{I_4}\nn
          &&+\frac{1}{2}C_{K_1K_3}L_{K_3K_2}B_{K_1}B_{K_2}.
\ee
There is no term $J_K^{I_1I_2}$ in the Hamiltonian. This term
was eliminated with help of the commutator (\ref{comm-rel}--ii).
Taking the time derivative and using the dynamics, one gets
\be\label{cons}
C_{K_1K}W^K_{I_1I_2}\Psi_{I_2}&=&2A_{I_1I}J^{II_2}_K\Psi_{I_2}.
\ee
This sort of consistency relation may be motivated by the
physical assumption, that there is no asymmetry in the
boson--fermion and fermion--boson interaction (actio equals
reactio, Newton). Furthermore a consistency consideration leads
to the above relation \cite{StuFauPfi}.

\subsection{\protect\label{SEC-6.1}Antisymmetric case}

We give the definitions of generating functionals for spinor
QED. We introduce $B$ sources and $\p^b$ duals
\be
[b_{K_1}, b_{K_2} ]_- = &0& = [ \p^b_{K_1}, \p^b_{K_2} ]_- \nn
{} [\p^b_{K_1},b_{K_2} ]_- &=& \delta_{K_1K_2}
\ee
analogous to the Schwinger sources. Therewith we define
\be
\vert {\cal A}(a,j,b)>_F &=& \sum_{m,n=0}^{\infty}
\frac{i^n}{n!m!} \tau_{nm}(I_1,\ldots,I_n,K_1,\ldots,K_m\mid a) \nn
&&
j_{I_1}\wedge\ldots\wedge j_{I_n} b_{K_1}\circ \ldots \circ b_{K_m}
\mid 0>_F \nn
\tau_{nm}(I_1,\ldots,I_n,K_1,\ldots,K_m\mid a)&:=&
<0\mid {\cal A}(\psi_{I_1}\ldots\psi_{I_n})
       {\cal S}(B_{K_1}\ldots B_{K_m})\mid a>
\ee
and
\be
\vert {\cal N}(a,j,b)>_F &=& \sum_{m,n=0}^{\infty}
\frac{i^n}{n!m!} \phi_{nm}(I_1,\ldots,I_n,K_1,\ldots,K_m\mid a)\nn
&&
j_{I_1}\dwedge\ldots\dwedge j_{I_n} b_{K_1}\circ \ldots \circ b_{K_m}
\mid 0>_F \nn
\phi_{nm}(I_1,\ldots,I_n,K_1,\ldots,K_m\mid a)&:=&
<0\mid {\cal N}(\psi_{I_1}\ldots\psi_{I_n})
       {\cal S}(B_{K_1}\ldots B_{K_m})\mid a>
\ee
where $\circ$ denotes the symmetric product of the $b_K$'s,
${\cal A}$ the antisymmetrizer, ${\cal S}$ the symmetrizer and
${\cal N}$ the normal-ordering and $\vert a>$ an energy
eigen-`state'. We assume furthermore, that the $j,\p$ and
$b,\p^b$ generators commute. We do not consider bosonic
normal-ordering. The field operators in the antisymmetric case
are given as
\be
\psi_I &=& \frac{1}{i}\p_I -
\frac{i}{2}A_{II^\prime}j_{I^\prime} \nn
\psi^{op}_I &=& \frac{1}{i}\p_I +
\frac{i}{2}A_{II^\prime}j_{I^\prime} \nn
B_K &=& \p^b_K - \frac{1}{2} C_{KK^\prime} b_{K^\prime} \nn
B^{op}_K &=& \p^b_K + \frac{1}{2} C_{KK^\prime} b_{K^\prime}.
\ee
Inserting these operators into (\ref{FKT-Heisenbergeq}) with the
Hamiltonian (\ref{HPsiB}) yields
\be\label{funct-1}
E_{0a}\vert{\bf A}(a,j,b)>&=&\big\{D_{I_1I_2}j_{I_1}\partial_{I_2}
+W^K_{I_1I_2}j_{I_1}\partial_{I_2}\partial^b_K
+L_{K_1K_2}b_{K_1}\partial^b_{K_2}\nn
&&+U_{I_1}^{I_2I_3I_4}j_{I_1}(\partial_{I_2}\partial_{I_3}\partial_{I_4}
  -\frac{1}{4}A_{I_3I_3^\prime}A_{I_2I_2^\prime}j_{I_2^\prime}j_{I_3^\prime}
  j_{I_4})\nn
&&+J_K^{I_1I_2}b_K(\partial_{I_1}\partial_{I_2}+\frac{1}{4}A_{I_1I_1^\prime}
  A_{I_2I_2^\prime}j_{I_1^\prime}j_{I_2^\prime})\big\}\vert{\bf A}
  (a,j,b)>,
\ee
where the $A_{I_nI_m}$ terms stem from the field quantization of
the fermions. Boson quantization terms would again occur, if
the $J_K^{I_1I_2}$ would be eliminated in the above mentioned
way. This result can be obtained from the normal-ordered
equation by setting $F\equiv 0$, so we treat that case in more
detail. 

\subsection{\protect\label{SEC-6.2}Normal-ordered case}

The (fermion) normal-ordered equation is obtained by using the
normal-ordered fermions
\be
\psi_I &=& \frac{1}{i}\p_I -
\frac{i}{2}A_{II^\prime}j_{I^\prime}
+i F_{II^\prime}j_{I^\prime} \nn
\psi^{op}_I &=& \frac{1}{i}\p_I +
\frac{i}{2}A_{II^\prime}j_{I^\prime} 
+i F_{II^\prime}j_{I^\prime}, 
\ee
the boson operators remain unchanged. Before we proceed, we give
the commutators of the fields, not displayed commutators vanish.
The bosons obey
\be\label{comm-funct}
{}[\partial^b_{K_1}\pm\frac{1}{2}C_{K_1K_1^\prime}b_{K_1^\prime},
\partial^b_{K_2}\pm\frac{1}{2}C_{K_2K_2^\prime}b_{K_2^\prime}]_-&=&
\mp C_{K_1K_2}\nn
{}[\partial^b_{K_1}\pm\frac{1}{2}C_{K_1K_1^\prime}b_{K_1^\prime},
\partial^b_{K_2}\mp\frac{1}{2}C_{K_2K_2^\prime}b_{K_2^\prime}]_-&=&0,
\ee
while we have for the normal ordered fermions
\be
\{\frac{1}{i}\partial_{I_1}\pm\frac{i}{2}A_{I_1I_1^\prime}j_{I_1^\prime}
+iF_{I_1I_1^\prime}j_{I_1^\prime},
\frac{1}{i}\partial_{I_2}\pm\frac{i}{2}A_{I_2I_2^\prime}j_{I_2^\prime}
+iF_{I_2I_2^\prime}j_{I_2^\prime}\}_+
&=& \pm A_{I_1I_2}\nn
\{\frac{1}{i}\partial_{I_1}\pm\frac{i}{2}A_{I_1I_1^\prime}j_{I_1^\prime}
+iF_{I_1I_1^\prime}j_{I_1^\prime},
\frac{1}{i}\partial_{I_2}\mp\frac{i}{2}A_{I_2I_2^\prime}j_{I_2^\prime}
+iF_{I_2I_2^\prime}j_{I_2^\prime}\}_+&=&0.
\ee
We calculate directly the (fermion) normal-ordered functional
Hamiltonian from $H[\Psi,B]^{\dwedge}$. Therefore we have to
build 
\be
H[j,b,\partial,\partial^b]^{\dwedge} &=& 
H[\frac{1}{i}\partial_I+\frac{i}{2}A_{II^\prime}j_{I^\prime} 
+iF_{II^\prime}j_{I^\prime},
 \partial^b_K+\frac{1}{2}C_{KK^\prime}b_{K^\prime}] \nn
&&-
H[\frac{1}{i}\partial_I-\frac{i}{2}A_{II^\prime}j_{I^\prime}
 +iF_{II^\prime}j_{I^\prime},
 \partial^b_K-\frac{1}{2}C_{KK^\prime}b_{K^\prime}].
\ee
We have according to (\ref{HPsiB}) four terms $T_i$. The $T_1$
term is of the same form as in the non-linear spinor field model
\be
T_1[j,\p]^{\dwedge} 
&=&
D_{I_1I_2}j_{I_1}\partial_{I_2}
-D_{I_1I_3}F_{I_3I_2}j_{I_1}j_{I_2},
\ee
where the renormalization was discussed in \cite{StuFauPfi}.
The term $T_2^{\dwedge}$ is given as
\be
T_2[j,\p]^{\dwedge}
&=&
\frac{1}{2}A_{I_1I_3}W^K_{I_3I_2}\Big[
(\partial^b_K+\frac{1}{2}C_{KK^\prime}b_{K^\prime})(\ldots
)(\ldots )\nn
&&-(\partial^b_K-\frac{1}{2}C_{KK^\prime}b_{K^\prime})(\ldots
)(\ldots )\Big]\nn
&=&
\frac{1}{2}A_{I_1I_3}W^K_{I_3I_2}\Big[\partial^b_K\big\{
(\ldots)(\ldots )-(\ldots)(\ldots )\big\}\nn
&&+\frac{1}{2}C_{KK^\prime}b_{K^\prime}\big\{
(\ldots)(\ldots )+(\ldots)(\ldots )\big\}
\Big].
\ee
The first term is just the same as before, while the second
yields 
\be
\mbox{2. brace}\quad
\{\}&=&-2\partial_{I_1}\partial_{I_2}-2\frac{1}{4}A_{I_1I_1^\prime}
A_{I_2I_2^\prime}j_{I_1^\prime}j_{I_2^\prime}\nn
&&+4F_{I_1I_1^\prime}j_{I_1^\prime}\partial_{I_2}
-2F_{I_1I_1^\prime}
F_{I_2I_2^\prime}j_{I_1^\prime}j_{I_2^\prime}.
\ee
We may use (\ref{cons}) to obtain
\be
\frac{1}{4}A_{I_1I_3}W^K_{I_3I_2}C_{KK^\prime}b_{K^\prime}
&=&
-\frac{1}{4}A_{I_1I_3}C_{K^\prime K}b_{K^\prime}W^K_{I_3I_2}\\
&=&
-\frac{1}{2}A_{I_1I_3}A_{I_3I_4}J^{I_4I_2}_{K^\prime}b_{K^\prime}\nn
&=&
-\frac{1}{2}J^{I_1I_2}_K b_K
\ee
and finally the result
\be
T_2[j,\p]^{\dwedge}
&=&W^K_{I_!I_2}\big[
j_{I_1}\partial_{I_2}-F_{I_2I_2^\prime}j_{I_1}j_{I_2^\prime}\big]
\partial^b_K\nn
&&+J_k^{I_1I_2}b_K\big[
\partial_{I_1}\partial_{I_2}
-2F_{I_1I_1^\prime}j_{I_1^\prime}\partial_{I_2}\nn
&&
+(F_{I_1I_1^\prime}F_{I_2I_2^\prime}+\frac{1}{4}
 A_{I_1I_1^\prime}A_{I_2I_2^\prime})j_{I_1^\prime}j_{I_2^\prime}
 \big].
\ee
The third term will be omitted, as it has the same structure as
$T_2$ in the non-linear spinor model. The last term runs along
the lines of $T_1$ and has not changed at all from
(\ref{funct-1}). Our result is then  
\be\label{funct-2}
E_{0a}\vert{\cal N}(a,j,b)>&=&\Big\{
 D_{I_1I_2}j_{I_1}\partial_{I_2}-Z_{I_1I_3}F_{I_3I_2}j_{I_1}j_{I_2}\nn
&&
+W^K_{I_!I_2}\big[
j_{I_1}\partial_{I_2}-F_{I_2I_2^\prime}j_{I_1}j_{I_2^\prime}\big]
\partial^b_K\nn
&&
+J_k^{I_1I_2}b_K\big[
 \partial_{I_1}\partial_{I_2}
 -2F_{I_1I_1^\prime}j_{I_1^\prime}\partial_{I_2}\nn
&&+(F_{I_1I_1^\prime}F_{I_2I_2^\prime}+\frac{1}{4}
    A_{I_1I_1^\prime}A_{I_2I_2^\prime})j_{I_1^\prime}j_{I_2^\prime}
    \big]\nn
&&
+U_{J}^{I_1I_2I_3}j_{J}
\big[
\partial_{I_1}\partial_{I_2}\partial_{I_3}
-3F_{I_3I_4}j_{I_4}\partial_{I_2}\partial_{I_1}\nn
&&
+(3F_{I_3I_4}F_{I_2I_5}+\frac{1}{4}
   A_{I_3I_4}A_{I_2I_5})j_{I_4}j_{I_5}\partial_{I_1}\nn
&&
-(F_{I_3I_4}F_{I_2I_5}F_{I_1I_6}+\frac{1}{4}
  A_{I_3I_4}A_{I_2I_5}A_{I_1I_6})j_{I_4}j_{I_5}j_{I_6}\big]\nn
&&
+L_{K_1K_2}b_{K_1}\partial^b_{K_2}
\Big\}\vert{\cal N}(a,j,b)>.
\ee
which coincides with (8.59) in \cite{StuBor} and (63) in
\cite{StuFauPfi}. We have succeeded in deriving the
normal-ordered functional equation in one single step by
Clifford algebraic considerations. No normal-ordering problems,
as a vertex normal ordering, was necessary. No singular
intermediate equation has occurred. And as a benefit, the
calculation was considerable short.

\section{\protect\label{SEC-7}Conclusion}

We have identified Clifford geometric algebras of multivectors
as the appropriate mathematical tool in modeling quantum field
theoretic generating functions. These algebras were studied in
section \ref{SEC-2}. Great emphasis was made on the connection
of the spaces underlying the Clifford and Grassmann algebras,
which induce the multivector structure in a Clifford algebra.
This structure is ubiquitously used in physics, but {\it not}\/
inherent in an ordinary Clifford algebra. However, if a Clifford
algebra is constructed from generators and relations, one has
implicitly imposed an ordering, which results in the explicit
multivector structure. Only the Clifford algebra of multivectors
can sytematically account for this situations and uncover the
role played by the propagator. The propagator or say
antisymmetric part of the contraction directly parameterizes the
vacua underlying the theory as was shown in \cite{Fau-vac}. 

The \ref{SEC-3}rd section develops an complete algebraic approach
to quantum field theoretic generating functionals by means of
Clifford algebras of multivectors. It was shown, that the field
operators can be seen as Clifford maps acting on the Grassmann
algebra build from the Schwinger source terms. The vacuum was
shown to be involved in the definition of matrix elements. This
result remains true for path integrals, which rely on the same
generating functionals. However, in the very compact notation of
path integrals no direct access to this structures is possible.

In section \ref{SEC-4} we gave the transition from operator
dynamics to generating functionals. This transition was
motivated by the Heisenberg equation. However, there is no
unique lifting of such an expression. To be able to reproduce
the results in literature, we chose the opposite action, not the
right action. Further investigations on that topic seem to be
necessary. 
The last sections \ref{SEC-5} and \ref{SEC-6} gave examples,
which showed the easiness and computational advantages of this
method. A non-linear spinor field was treated in the
antisymmetric and normal-ordered case. The same was done for
spinor QED as an example for a boson-fermion coupling theory.
The most remarkable result was, that it was possible to derive
normal-ordered generating functionals directly, that is in one
single step, from the operator dynamics. Furthermore, this
showed that the algebraic theory has full control over the
involved orderings and thereby over the involved vacuum states.
This was investigated in detail in \cite{Fau-vac}. The method
was already applied to Dirac theory with some success
\cite{Fau-ach}. 

\section{Acknowledgements}

Two long e-mail letters from Prof. Z. Oziewicz, which improved
semantic an clearness of section \ref{SEC-2}, were a valuable
help in preparing this work. The support of Mrs. Ursula Wieland
is gratefully acknowledged.  

\end{document}